\documentclass[aps,prl,twocolumn,groupedaddress]{revtex4}

\begin{document}

\preprint{TIT/HEP--521}
\preprint{hep-th/0404198}

\title{
Construction of 
Non-Abelian Walls 
and Their 
Complete 
Moduli Space 
}

\author{Youichi~Isozumi}
\email[]{isozumi@th.phys.titech.ac.jp}
\author{Muneto~Nitta}
\email[]{nitta@th.phys.titech.ac.jp}
\author{Keisuke~Ohashi}
\email[]{keisuke@th.phys.titech.ac.jp}
\author{Norisuke~Sakai}
\email[]{nsakai@th.phys.titech.ac.jp}
\affiliation{ Department of Physics, Tokyo Institute of 
Technology,
Tokyo 152-8551, JAPAN }


\date{\today}

\begin{abstract}
We present a systematic method to construct exactly 
all Bogomol'nyi-Prasad-Sommerfield 
(BPS) 
multi-wall solutions in supersymmetric (SUSY) 
$U(N_{\rm C})$ gauge theories 
in five dimensions 
with 
$N_{\rm F}$ hypermultiplets in the fundamental 
representation for infinite gauge coupling. 
The moduli space of these non-Abelian walls is found to be 
the complex Grassmann manifold 
${SU(N_{\rm F}) \over 
SU(N_{\rm C})\times SU(N_{\rm F}-{N}_{\rm C}) \times U(1)}$ 
endowed with a deformed metric. 
\end{abstract}


\maketitle
In constructing unified theories with extra 
dimensions~\cite{braneworld}
it is crucial to obtain topological defects and localization 
of massless or nearly massless modes on the defect. 
Walls in five-dimensional theories are 
the simplest of the topological defects leading to 
the four-dimensional world volume. 
In constructing topological defects, 
SUSY 
theories are helpful, since partial preservation of 
SUSY automatically gives a solution of equations of 
motion~\cite{WittenOlive}. These states are called 
BPS
states. 
The minimum number of supercharges in five-dimensions 
is eight. 
Wall solutions in the $U(1)$ gauge theory, 
which can be called {\it Abelian walls}, 
have been discussed with 8 SUSY~\cite{DW1}--\cite{IOS1}. 
Walls in non-Abelian gauge theories have been considered 
in a special circumstance recently\cite{SY2}. 
Non-Abelian solitons admitting ample moduli space structure 
were discovered already except for walls and  
beautiful methods are available for the construction 
of instantons, monopoles and vortices~\cite{ADHM}.
In this Letter, we give a systematic method 
to construct walls in non-Abelian gauge theories, 
called {\it non-Abelian walls}, 
with the gauge group 
$U(N_{\rm C})$ 
and $N_{\rm F} (>N_{\rm C})$ copies of hypermultiplets in 
the fundamental representation. 
We also find the complete moduli space for non-Abelian 
walls which fills the last gap 
in soliton moduli spaces in 
the gauge-Higgs system.

We shall denote the gauge group by the uppercase suffix C, 
and the flavor group by 
F. 
The $U(N_{\rm C})$ vector multiplets 
contain gauge fields $W_M$, and a real 
scalar field $\Sigma$, 
which are in the adjoint representation of 
$U(N_{\rm C})$. 
We use an $N_{\rm C}\times N_{\rm C}$ matrix notation for 
these component fields, like $\Sigma=\Sigma^I T_I$. 
Here we have denoted the Hermitian generators in the 
Lie algebra by $T^I, \,(I =0,1,2,\cdots, N_{\rm C}^2-1)$, 
satisfying the normalization condition;
${\rm Tr}(T_I T_J) = {1\over 2}\delta_{IJ}$,  
where $T_0$ is the generator of the factor $U(1)$ 
gauge group.
The  $U(1)$ part of vector multiplets allows 
the Fayet-Illiopoulos (FI) term which gives rise to 
discrete vacua once mass terms for hypermultiplets 
are introduced. 
Dynamical bosons of hypermultiplets are $SU(2)_R$ doublet of 
complex scalar quark fields $H^{irA}$. 
We denote space-time indices by 
$M,N, \cdots=0,1,2,3,4$, 
and $SU(2)_R$ 
doublet indices by $i$. 
The color indices $r,s,\cdots$ run over 
$1,2,\cdots, N_{\rm C}$, 
whereas $A,B, \cdots =1,2,\cdots, N_{\rm F}$ stand 
for flavor indices. 
It is convenient to combine the $N_{\rm F}$ 
hypermultiplets in the fundamental representation 
into an $N_{\rm C}\times N_{\rm F}$ matrix 
$H^i$ with components $(H^i)^{rA}\equiv H^{irA}$. 

We shall consider a model with minimal 
kinetic terms for vector and hypermultiplets. 
The 8 SUSY 
allow only a few parameters in our model: 
the masses of the $A$-th hypermultiplet $m_A$, 
the $SU(2)_R$ triplet of FI parameters $c_a,\,(a=1,2,3)$ for the  $U(1)$  
vector multiplet, and a 
gauge coupling constant $g$ for the $U(N_{\rm C})$ 
gauge group. 
Different gauge couplings 
for $U(1)$ and 
$SU(N_{\rm C})$ factors can easily be 
incorporated, 
but the difference becomes 
irrelevant for infinite gauge coupling 
which we will be most interested in. 
After eliminating the auxiliary fields, 
the bosonic part of our Lagrangian reads 
\begin{eqnarray}
{\cal L} &\!\!\!=&\!\!\! 
-\frac{1}{2g^2}{\rm Tr}(F_{MN}(W)F^{MN}(W))
 +\frac{1}{g^2}{\rm Tr}({\cal D}^M \Sigma {\cal D}_M \Sigma) \nonumber\\
&\!\!\!&\!\!\!
+ ({\cal D}_M H_{irA})^\dagger  {\cal D}^M H^{irA} -V,
\label{fundamental-Lag2}
\end{eqnarray}
where the scalar potential $V$ is given by
\begin{eqnarray}
V&=& 
\frac{g^2}{4}{\rm Tr}\Big[
\left((\sigma _a)^j{}_iH^iH_j^\dagger 
-c_a{\bf 1}_{N_{\rm C}}
\right)^2\Big] \nonumber\\
&&\qquad {}+H_{irA}^\dagger [(\Sigma -m_A)^2]^r{}_s H^{isA}.
\label{eq:potential}
\end{eqnarray}
Here a sum over repeated indices is implied, 
covariant derivatives are defined by 
${\cal D}_M H^{irA}
=(\partial_M \delta_s^r + iW_M^I (T_I)^r{}_s)H^{isA}$, 
${\cal D}_M \Sigma = \partial_M \Sigma + i[ W_M , \Sigma ]$, 
the gauge field strength is defined by
$F_{MN}(W)=-i[{\cal D}_M , {\cal D}_N]
$. 
Our convention of metric is 
$\eta_{MN}={\rm diag}(+1,-1,-1,-1,-1)$. 
In this Letter, we assume non-degenerate mass parameters 
$m_A$, which we arrange $m_A > m_{A+1}$ for all $A$.  
Our results should be 
valid for the degenerate mass case also, 
except for subtleties associated with global symmetry. 
Since $U(1)_{\rm F}$ corresponding to a common phase is 
gauged, 
the flavor symmetry reduces to 
$U(1)_{\rm F}^{N_{\rm F}-1}$.
The $SU(2)_R$ symmetry allows us to 
choose the FI parameters to lie in the third direction 
without loss of generality as $c_a =(0,\ 0,\ c)$ with $c>0$. 

Let us discuss the vacuum structure of this model.
Since we assume non-degenerate masses for hypermultiplets, 
we find that only one flavor $A=A_r$ 
($A_r\not=A_s$, for $r\not=s$) 
can be non-vanishing 
for each color component $r$ of hypermultiplet scalars 
$H^{irA}$ with 
\begin{eqnarray}
 H^{1rA}=\sqrt{c}\,\delta ^{A_r}{}_A,\quad H^{2rA}=0.
 \label{eq:hyper-vacuum}
\end{eqnarray}
Here we used global gauge transformations 
to eliminate possible phase factors. 
This is called the color-flavor locking vacuum. 
The vector multiplet scalars $\Sigma$ are 
determined as
\begin{eqnarray}
\Sigma ={\rm diag}(m_{A_1},\,m_{A_2},\,\cdots,\,
m_{A_{N_{\rm C}}}).
\end{eqnarray}
We denote a SUSY vacuum specified by a set of 
non-vanishing hypermultiplet 
scalars with the flavor $\{A_r\}$ for each color 
component $r$ as 
$
 \langle A_1 A_2 \cdots A_{N_{\rm C}}\rangle $. 
Since global gauge transformations can exchange flavors 
$A_i$ and $A_j$ for the color component $i$ and $j$, 
respectively, 
the ordering of the flavors $A_1, \cdots, A_{N_{\rm C}}$ 
does not matter in considering only vacua: 
$\langle 1,2,3 \rangle
=\langle 2,1,3 \rangle
$. 
Thus a number of SUSY 
vacua is given by $
N_{\rm F}!/((N_{\rm F}-N_{\rm C})!N_{\rm C}!)\equiv 
{}_{N_{\rm F}}C_{N_{\rm C}}$ 
and we usually take $A_1<A_2<\cdots<A_{N_{\rm C}}$. 
(Multi-)walls are classified by topological sectors 
that are defined by 
giving two vacua at $y=\pm\infty$. 

Let us obtain the BPS equations for domain walls 
interpolating between two SUSY vacua. 
We require for wall solutions that 
all fields depend only on the coordinate of 
the extra dimension $y\equiv x^4$. 
We also assume the Poincar\'e invariance 
on the four-dimensional world volume of the wall, 
implying $W_M=0$ for the indices $M\not=y$.
Note that $W_y$ need not vanish. 
We demand that half of SUSY defined by $\gamma ^4\varepsilon ^i
=-i(\sigma ^3)^i{}_j\varepsilon ^j$ 
to be conserved~\cite{IOS1}.  
Requiring the SUSY transformation of fermions to vanish 
along the above SUSY directions, 
we find the following BPS equations for domain walls 
in the matrix notation 
\begin{eqnarray}
{\cal D}_y \Sigma &=& 
{g^2\over 2}\left(c{\bf 1}_{N_{\rm C}}-H^1H^1{}^\dagger 
+H^2H^2{}^\dagger \right), 
\label{BPSeq-Sigma}\\
0&=& g^2 H^1H^2{}^\dagger ,\label{BPSeq-Y12}\\
{\cal D}_y H^1 &=& -\Sigma H^1 + H^1 M,\quad 
{\cal D}_y H^2 = \Sigma H^2 -H^2 M,\label{BPSeq-H}\qquad 
\end{eqnarray}
where we have used the $N_{\rm F}\times N_{\rm F}$ Hermitian 
mass matrix $M$ defined by  
 $(M)^A{}_B\equiv m_A\delta ^A{}_B$.

If a wall configuration approaches 
a SUSY vacuum 
$\langle A_1A_2\cdots A_{N_{\rm C}}\rangle $ 
at 
$y=+\infty  $, and 
$\langle B_1B_2\cdots B_{N_{\rm C}}\rangle $ 
at 
$y=-\infty  $, 
the topological sector of the configuration 
is labeled by 
$\langle A_1A_2\cdots A_{N_{\rm C}}\rangle 
\leftarrow 
\langle B_1B_2\cdots B_{N_{\rm C}}\rangle $. 
By either performing 
the Bogomol'nyi completion of the energy density 
$\cal E$ or applying the BPS equations, we obtain the 
bound for the energy of the configuration 
\begin{eqnarray}
\int^{+\infty}_{-\infty}{\cal E}dy
\geq c \Big[ {\rm Tr}(\Sigma )\Big]^{+\infty}_{-\infty}
=c\,\left(\sum_{k=1}^{N_{\rm C}}m_{A_k}
-\sum_{k=1}^{N_{\rm C}}m_{B_k}\right). 
\;
\end{eqnarray}
BPS walls saturate the bound. 

Let us construct solutions for BPS 
Eqs.~(\ref{BPSeq-Sigma})--(\ref{BPSeq-H}). 
To this end, it is convenient to introduce an 
$N_{\rm C}\times N_{\rm C}$ invertible 
complex matrix function $S(y)\in GL(N_{\rm C},\mathbf{C})$ defined by\footnote{
Our matrix function $S$ is a non-Abelian generalization 
of a  complex function 
$\psi={\rm log} S$ introduced to solve the BPS equation for 
Abelian walls~\cite{Tong,IOS1}. 
}
\begin{eqnarray}
\Sigma + iW_y \equiv S^{-1}\partial_y S.
\label{def-S}
\end{eqnarray}
Note that the above 
differential equation determines the matrix 
function $S$ except for  
$N_{\rm C}^2$ complex integration constants which 
cause an ambiguity
for $S$. 
Without any assumption, 
the BPS eqs.~(\ref{BPSeq-Y12}) and 
(\ref{BPSeq-H}) dictate 
\begin{eqnarray}
 H^1 = S^{-1}H_0 e^{My}, 
 \quad H^2 = 0 .
\label{sol-H}
\end{eqnarray}
Here $H_0$ is an arbitrary complex constant 
$N_{\rm C} \times N_{\rm F}$ 
matrix which we call the ``moduli matrix''.
We will postpone detailed proof (including $H^2=0$) 
in a subsequent paper. 
The remaining BPS eq.~(\ref{BPSeq-Sigma}) 
for the vector multiplets 
can be written in terms of the matrix $S$ and the moduli 
matrix $H_0$.   
Eq.~(\ref{def-S}) implies that 
the gauge transformations on the original 
fields $\Sigma ,\,W_y,\,H^1$
\begin{eqnarray}
H^1  &\rightarrow&  
H^1{}'\! =\! U H^1,\nonumber \\
\Sigma +iW_y &\rightarrow&
\Sigma' +iW_y' \!
=\! U\left(\Sigma +iW_y\right)U^\dagger 
+ U\partial_y U^\dagger \quad \,
\end{eqnarray}
can be obtained by a right-multiplication of 
a unitary matrix $U^\dagger$ on $S$:  
\begin{eqnarray}
 S\,\rightarrow \,S'=SU^\dagger ,\quad U^\dagger U=1
\end{eqnarray}
without causing any transformations on the moduli 
matrix $H_0$. 
Therefore we obtain 
gauge invariant quantity $\Omega $ out of $S$ 
defined by 
\begin{eqnarray}
 \Omega \equiv SS^\dagger .\label{def-Omega}
\end{eqnarray}
Together with the gauge invariant moduli matrix $H_0$, 
the BPS eq.~(\ref{BPSeq-Sigma}) 
can be rewritten in the following 
gauge invariant form 
\begin{eqnarray}
\partial_y^2 \Omega -\partial_y \Omega \Omega^{-1} \partial_y \Omega = g^2 
\left(c\, \Omega - H_0 \,e^{2My} H_0{}^\dagger 
\right).\label{diff-eq-S}
\end{eqnarray}
With a suitable gauge choice, we obtain uniquely the 
$N_{\rm C}\times N_{\rm C}$ complex matrix $S$ 
from the $N_{\rm C}\times N_{\rm C}$ Hermitian 
matrix $\Omega $. 
Therefore, once a solution of $\Omega $ for  
Eq.(\ref{diff-eq-S}) with a given moduli matrix $H_0$ is 
obtained, 
the matrix $S$ can be determined and then, 
all the quantities, $\Sigma ,\,W_y$ and $H^1$ 
are obtained by Eqs.~(\ref{def-S}) and (\ref{sol-H}). 
We find by explicit examples that gauge field 
$W_y$ and/or $\Sigma$ are 
nontrivial unlike Abelian walls. 

Given the boundary conditions at both infinities 
$y=\pm \infty$, the differential eq.~(\ref{diff-eq-S}) 
is expected to give a solution without further 
integration constants. 
Therefore the moduli matrix $H_0$ alone 
should describe the entire moduli space of walls. 
Eq.~(\ref{diff-eq-S}) is, however, difficult to solve 
explicitly for finite 
gauge couplings $g$. 
We consider, therefore, the case of the infinite 
gauge coupling 
($g^2\rightarrow \infty $), where 
Eq.~(\ref{diff-eq-S}) for the gauge invariant $\Omega$ 
reduces to an algebraic equation, given by 
\begin{eqnarray}
 \Omega_{g \to \infty} 
 = (SS^\dagger)_{g \to \infty}  
 = c^{-1}H_0 e^{2My}H_0^\dagger. 
  \label{SS-H0}
\end{eqnarray}
Therefore 
we can explicitly construct wall solutions 
in the infinite gauge coupling
without solving the 
differential 
equation for $\Omega$. 
This explicit solution shows clearly that the moduli space 
is fully covered by our moduli matrix $H_0$. 
In this limit 
our model reduces to 
a hyper-K\"ahler (HK) nonlinear sigma model 
(NLSM) whose target space is 
the cotangent bundle over the complex 
Grassmann manifold 
$T^* [{ SU(N_{\rm F}) \over SU(N_{\rm C}) 
\times SU(N_{\rm F} - N_{\rm C}) \times U(1)}]$~\cite{ANS}. 
For this NLSM, 
our construction exhausts all possible 
BPS wall solutions. 
The NLSM has been known to be dual under 
$N_{\rm C} \leftrightarrow N_{\rm F}-N_{\rm C}$ with 
$N_{\rm F}$ fixed. 
We find duality transformations of moduli matrix $H_0$ 
explicitly. 
For the non-Abelian gauge theory in 
Eqs.(\ref{fundamental-Lag2}) and (\ref{eq:potential}), 
it is likely that one 
needs to consider finite gauge couplings, 
especially if one is interested in quantum effects. 
The BPS domain walls in theories with 
8 SUSY were first obtained in 
HK NLSMs~\cite{DW1}. 
They have been the only known examples for 8 SUSY models 
until exact wall solutions 
at finite gauge coupling were found 
recently~\cite{KS,IOS1}. 
In \cite{IOS1} 
we have constructed exact wall solutions 
for finite gauge couplings in the case of 
$N_{\rm C}=1$ and $N_{\rm F}=3$ 
to find that their qualitative behavior 
is the same as the infinite gauge coupling cases 
found in 
\cite{Tong}. 
We expect that the moduli space of walls at finite 
gauge couplings should be qualitatively the same as 
that at infinite gauge coupling. 
In the rest of this Letter we examine moduli matrix 
$H_0$ irrespective of finite or infinite gauge coupling. 

From Eqs.~(\ref{def-S}) and 
(\ref{sol-H}), 
we find that 
the same original fields $\Sigma ,\,W_y, H^1$ 
given by a set 
of matrix function $S$ and constant 
moduli matrix $H_0$ 
are described by 
another set $(S', H_0{}')$ 
 transformed by $V\in GL(N_{\rm C},\mathbf{C})$ 
\begin{eqnarray}
 S\rightarrow S' = VS,\qquad 
H_0 \rightarrow H_0{}'=VH_0. 
\label{art-sym}
\end{eqnarray} 
We call this global 
``world-volume symmetry'', which comes from 
the $N_{\rm C}^2$ integration constants in solving 
(\ref{def-S}). 
This transformation $V$ defines an equivalence class 
among sets of matrix function $S$ and moduli matrix 
$H_0$. 
We thus find the moduli space for (multi-)wall solutions 
(without specifying boundary conditions)
denoted by ${\cal M}_{N_{\rm F},N_{\rm C}}$
is 
 the complex Grassmann manifold:
\begin{eqnarray}
 {\cal M}_{N_{\rm F},N_{\rm C}}
 &=
 &\{H_0 | H_0 \sim V H_0, V \in GL(N_{\rm C},\mathbf{C})\} \nonumber\\
\equiv
G_{N_{\rm F},N_{\rm C}}
 &\simeq & {SU(N_{\rm F}) \over 
 SU(N_{\rm C}) \times SU(N_{\rm F}-N_{\rm C}) 
 \times U(1)}\,,\qquad 
  \label{Gr}
\end{eqnarray} 
whose complex dimension is given by 
$
N_{\rm C} (N_{\rm F}-N_{\rm C})$. 
This is a {\it compact} (closed) set. 
On the other hand, for instance,  
scattering of two Abelian walls is described by 
a NLSM on a {\it non-compact} moduli space~\cite{Tong,IOS1}. 
We also find similar non-compact moduli by an explicit 
analysis of multiple non-Abelian walls. 
These two facts can be consistently understood, if 
we note that  
the moduli space 
${\cal M}_{N_{\rm F},N_{\rm C}}$ 
includes {\it all topological sectors} 
determined by the different boundary conditions 
as we show in the rest
of this Letter.

The moduli matrix $H_0$ contains complete data of walls 
including boundary conditions, number of walls, wall 
position, etc. 
Boundary conditions at $y=\pm \infty$ 
are most conveniently read by 
the following fixing of world-volume symmetry 
(\ref{art-sym}) : 
\begin{widetext}
 \begin{eqnarray}
&&\hspace{5em}A_1\hspace{2.0em}A_2
\hspace{6.5em}B_1\hspace{3.7em}B_2\nonumber\\
 H_0&=&\sqrt{c}\left(
\begin{array}{cccccccccccc}
\cdots 0&1&*\cdots &*&\cdots  & &\cdots *&e^{v_1} 
&0\cdots     \\
 & &\cdots 0&1&*\cdots & &\cdots  &    &\cdots*&e^{v_2}
 &0\cdots\\
 & &\vdots  & &        & &\vdots  &      &                 
 &&& \\
 & &  & &\cdots 0&1&*\cdots &\cdots *&e^{v_{N_{\rm C}}}
 &0\cdots\\
\end{array}\right),
\label{standard-form}\\
&&\hspace{11.7em}A_{N_{\rm C}}\hspace{4.5em} B_{N_{\rm C}} 
\nonumber
\end{eqnarray}
\end{widetext}
where all elements in the $r$-th row 
before the $A_r$-th flavor are eliminated, the $A_r$-th 
flavor is normalized to be unity, 
and the last non-vanishing element $e^{v_r}$ 
($v_r \in {\bf C}$) in the 
$r$-th row resides in the $B_r$-th flavor. 
We can choose these flavors $A_r, B_r$ to be ordered as 
\begin{equation}
1\leq A_1<A_2<\cdots<A_{N_{\rm C}}\leq N_{\rm F},
\label{eq:echelon-ordering}
\qquad 
\end{equation}
\vspace{-2.5em}
\begin{equation}
 A_r\leq B_r,
\qquad  B_r\not=B_s,\quad  {\rm for~} r\not=s .
\label{eq:ordering-AB}
\end{equation}
When the set of flavors 
$\{B_r\}$ are not ordered like $\{A_r\}$ in 
Eq.~(\ref{eq:echelon-ordering}), 
we must eliminate some more elements 
to remove the redundancy. 
This can be done in a well-defined procedure. 
We call the fixing (\ref{standard-form}) a 
``standard form''. 
Since this fixing 
of the symmetry (\ref{art-sym}) 
is unique, 
any moduli matrix in the standard form has one-to-one 
correspondence with a point in the moduli space. 
If the moduli matrix happens to be 
$H_0^{rA}=\sqrt{c} \delta^{A_r}{}_A$, 
Eqs.~(\ref{sol-H}) and (\ref{SS-H0}) imply the vacuum 
$\langle A_1A_2\cdots A_{N_{\rm C}}\rangle$ : 
$H^{1rA}(y)=\sqrt{c} \delta^{A_r}{}_A$. 
Note that the solution 
$H^1$ in Eq.~(\ref{sol-H}) implies the transformation of 
the moduli matrix, $H_0\rightarrow H_0e^{My_0}$, 
under a translation $y\rightarrow y+y_0$.  
Since the world-volume symmetry 
(\ref{art-sym}) 
allows us to 
multiply the matrix 
$(V)^r{}_s=e^{-m_{A_r}y_0}\delta ^r{}_s$ 
from the left of $H_0$, 
the standard 
form (\ref{standard-form}) and the ordering of masses 
imply that 
the matrix $(VH_0e^{My_0})^{rA}$ 
remains finite when taking 
the limit $y_0\rightarrow \infty $ to give 
$\sqrt{c}\delta ^{A_r}{}_A$. 
Thus the configuration reduces to the 
vacuum labelled by 
$\langle A_1A_2\cdots A_{N_{\rm C}}\rangle $. 
Similarly, with another matrix 
$(V)^r{}_s=e^{-m_{B_r}y_0-v_r}
\delta^r{}_s
$, we obtain 
$(VH_0e^{My_0})^{rA} \rightarrow \sqrt{c}\delta ^{B_r}{}_A$ 
in the limit of $y_0\rightarrow -\infty $.
Therefore 
the multi-wall configuration 
described by 
the standard form 
(\ref{standard-form}) 
belongs to the 
{\it topological sector labeled by} 
$\langle A_1A_2\cdots A_{N_{\rm C}}\rangle $
$\leftarrow\langle B_1B_2\cdots B_{N_{\rm C}}\rangle$. 

A topological sector consists of all permutations 
of the vacuum labels $B_1, B_2,\cdots, B_{N_{\rm C}}$ 
at $y=-\infty $. 
If the label 
$\langle B_1B_2\cdots B_{N_{\rm C}}\rangle $ 
happens to be 
ordered, $B_1 < B_2 < \cdots <  B_{N_{\rm C}}$, 
then the moduli matrix $H_0$ covers generic points of the 
topological sector. 
Hence the real dimension of the topological sector 
is given by 
$
 2\left(\sum_{i=1}^{N_{\rm C}}B_i
-\sum_{i=1}^{N_{\rm C}}A_i\right)$.
Half of these moduli parameters represent wall 
positions and the rest are (quasi-)Nambu-Goldstone 
modes of internal symmetry. 
The topological sector with the largest 
dimension is labelled by 
$\langle 1,2,\cdots,N_{\rm C}\rangle 
\leftarrow \langle N_{\rm F}-N_{\rm C}+1,
\cdots,N_{\rm F}-1,N_{\rm F} \rangle $.
If the label 
$\langle B_1B_2\cdots B_{N_{\rm C}}\rangle $ 
is not ordered, 
$H_0$ has smaller dimensions as is described below 
Eq.(\ref{eq:ordering-AB}). 
We can understand this fact by noting that 
some walls are compressed each other to become 
a single ``compressed wall''. 

By the above observation, 
we find that the Grassmann manifold 
is decomposed into 
\begin{eqnarray}
 {\cal M}_{N_{\rm F},N_{\rm C}} 
 = \sum_{\rm BPS} 
 {\cal M}^{\langle A_1A_2\cdots A_{N_{\rm C}}\rangle 
 \leftarrow \langle B_1B_2\cdots 
 B_{N_{\rm C}}\rangle }_{N_{\rm F},N_{\rm C}} ,\;
\end{eqnarray}
where ${\cal M}^{\langle A_1A_2\cdots A_{N_{\rm C}}\rangle \leftarrow  
\langle B_1B_2 \cdots 
B_{N_{\rm C}}\rangle }_{N_{\rm F},N_{\rm C}}$
denotes the moduli subspace 
of BPS (multi-)wall solutions for the topological sector of 
$\langle A_1A_2\cdots A_{N_{\rm C}}\rangle$
$\leftarrow \langle B_1B_2 \cdots B_{N_{\rm C}}\rangle $.
Note that it also includes the vacuum states with no walls 
$\langle A_1A_2\cdots A_{N_{\rm C}} \rangle$
$\leftarrow$ $\langle A_1A_2 \cdots A_{N_{\rm C}} \rangle$ 
which correspond to 
${}_{N_{\rm F}}C_{N_{\rm C}}$ points 
on the moduli space.
Although each sector (except for vacuum states) 
is in general an open set,
the total space is compact. 
We call 
${\cal M}_{N_{\rm F},N_{\rm C}}$ as 
the ``total moduli space''. 
This fact is in 
interesting contrast 
to cases of other solitons like instantons, vortices 
and monopoles, since 
the dimension of the total moduli spaces is infinite 
in the latter cases. 

Effective Lagrangians on walls can be obtained 
by promoting  the moduli parameters to fields 
on the world-volume of walls
\cite{Ma}. 
The world-volume symmetry 
(\ref{art-sym}) naturally becomes 
a local gauge symmetry. 
Denoting the moduli fields by $\phi$ in $H_0(\phi)$, 
we obtain 
the K\"ahler metric on the total moduli space. 
By using explicit solutions for infinite gauge coupling, 
its K\"ahler potential is given by 
\begin{eqnarray}
 K(\phi ,\phi^*) = c \int d y \,
 \log\left[{\rm det}\Omega (\phi ,\phi^*,y)\right]
  \,,
  \label{eq:K-potential}
\end{eqnarray}
which is expected to be 
valid for finite coupling too. 
The metric (\ref{eq:K-potential}) is not symmetric 
under $SU(N_{\rm F})$ 
but admits an isometry $U(1)^{N_{\rm F}-1}$. 
Therefore the total moduli space is
a deformed Grassmann manifold.

The total moduli space $G_{N_{\rm F},N_{\rm C}}$ 
is a special Lagrangian submanifold 
of the Higgs branch 
of vacua $T^* G_{N_{\rm F},N_{\rm C}}$ of this theory. 
We anticipate that this is always true for arbitrary 
gauge group with arbitrary matter contents.

\begin{acknowledgments}
The authors thank David Tong for useful discussion 
and Masato Arai for a collaboration 
at an early stage. 
This work is supported in part by Grant-in-Aid for Scientific 
Research from the Ministry of Education, Culture, Sports, 
Science and Technology, Japan No.13640269 
and 16028203 (NS and MN). 
K.O.~and M.N.~are 
supported in part by 
JSPS and 
Y.I.~a 21st Century COE Program at 
Tokyo Tech ``Nanometer-Scale Quantum Physics''. 
\end{acknowledgments}

\bibliography{}

\begin{thebibliography}{100}


\bibitem{braneworld}
    P.~Horava and E.~Witten, 
     {Nucl.\ Phys.\ } {\bf B460}, 506 (1996);  
 N.~Arkani-Hamed, S.~Dimopoulos and G.~Dvali, 
             {Phys Lett.} {\bf B429}, 263 (1998);  
             I.~Antoniadis, N.~Arkani-Hamed, S.~Dimopoulos 
             and G.~Dvali, 
             {Phys.\ Lett.} {\bf B436}, 257 (1998);   
             L.~Randall and R.~Sundrum, 
             {Phys.\ Rev.\ Lett.} 
             {\bf 83}, 3370 (1999); 
             {Phys.\ Rev.\ Lett.} {\bf 83}, 4690 (1999).  

 \bibitem{WittenOlive} E.~Witten and D.~Olive, 
             {Phys.\ Lett.\ } {\bf B78}, 97 (1978) .

\bibitem{DW1}
E.~Abraham and P.~K.~Townsend, 
             {Phys.\ Lett.} {\bf B 291}, 85 (1992);
J.~P.~Gauntlett, R.~Portugues, D.~Tong, and P.K.~Townsend, 
               {Phys.\ Rev.} {\bf D63}, 085002 (2001); 
J.~P.~Gauntlett, D.~Tong, and P.~K.~Townsend, 
               {Phys.\ Rev. } {\bf D63}, 085001 (2001);  
               {Phys.\ Rev. }{\bf D64}, 025010 (2001); 
D.~Tong, 
{JHEP }{\bf 0304}, 031 (2003); 
M.~Arai, M.~Naganuma, M.~Nitta and N.~Sakai, 
{Nucl.\ Phys. }{\bf B652}, 35 (2003); 
in Garden of Quanta - In honor of Hiroshi Ezawa, 
World Scientific Publishing Co. Pte. Ltd., Singapore, 
hep-th/0302028;
M. Arai, E.~Ivanov and J.~Niederle, 
{Nucl.\ Phys. }{\bf B680}, 23 (2004).  


\bibitem{Tong}  D.~Tong, 
               {Phys.\ Rev.} {\bf D66}, 025013 (2002). 

\bibitem{DW2}
K.~S.~M.~Lee, 
             {Phys.\ Rev.\ }{\bf D67}, 045009 (2003);  
M.~Shifman and A.~Yung
{Phys.\ Rev.\ }{\bf D67}, 125007 (2003). 

\bibitem{KS}
K.~Kakimoto and N.~Sakai, 
{Phys.\ Rev.\ }{\bf D68}, 065005 (2003). 

\bibitem{IOS1} Y.~Isozumi, K.~Ohashi, and N.~Sakai, 
{JHEP }{\bf 11}, 060 (2003); 
 {JHEP }{\bf 11}, 061 (2003).


\bibitem{SY2}
M.~Shifman and A.~Yung,
hep-th/0312257.


\bibitem{ADHM} 
M.~F.~Atiyah, N.~J.~Hitchin, V.~G.~Drinfeld and Yu.~I.~Manin,
{Phys.\ Lett.} {\bf A65}, 185 (1978);  
W.~Nahm, 
{Phys.\ Lett.} {\bf B90}, 413 (1980);
A.~Hanany and D.~Tong, 
{JHEP }{\bf 0307}, 037 (2003).

\bibitem{ANS}   M.~Arai, M.~Nitta, and N.~Sakai, 
hep-th/0307274.


\bibitem{Ma}
N.~S.~Manton, 
{Phys.\ Lett. }{\bf B110}, 54 (1982).


\end{thebibliography}
\newcommand{\J}[4]{{\sl #1} {\bf #2} (#3) #4}
\newcommand{\andJ}[3]{{\bf #1} (#2) #3}
\newcommand{\AP}{Ann.\ Phys.\ (N.Y.)}
\newcommand{\MPL}{Mod.\ Phys.\ Lett.}
\newcommand{\NP}{Nucl.\ Phys.}
\newcommand{\PL}{Phys.\ Lett.}
\newcommand{\PR}{ Phys.\ Rev.}
\newcommand{\PRL}{Phys.\ Rev.\ Lett.}
\newcommand{\PTP}{Prog.\ Theor.\ Phys.}
\newcommand{\hep}[1]{{\tt hep-th/{#1}}}

\end{document}